\documentclass[twocolumn]{revtex4}
\usepackage{amsmath}
\usepackage{amssymb}
\usepackage{color}
\usepackage{ulem}
\usepackage{graphicx}

\newcommand{\ra}{\rangle}
\newcommand{\la}{\langle}

\begin{document}

\title{Imaginary-Time Path-Integral in Bulk Space from the Holographic Principle}
%%Imaginary-time path-integral in bulk space from the holographic principle}

\author{Eiji Konishi}
\email{konishi.eiji.27c@kyoto-u.jp}
\address{Graduate School of Human and Environmental Studies, Kyoto University, Kyoto 606-8501, Japan}

\date{\today}

\begin{abstract}
In the three-dimensional anti-de Sitter spacetime/two-dimensional conformal field theory correspondence, we derive the imaginary-time path-integral of a non-relativistic particle in the anti-de Sitter bulk space, which is dual to the ground state, from the holographic principle.
This derivation is based on
(i) the author's previous argument that the holographic principle asserts that the anti-de Sitter bulk space as a holographic tensor network after classicalization has as many stochastic classicalized spin degrees of freedom as there are sites and
(ii) the reinterpretation of the Euclidean action of a free particle as the action of classicalized spins.
\end{abstract}

\maketitle

\section{Introduction}
The holographic principle is the duality between gravity in a bulk spacetime and quantum field theory in the one-codimensional boundary spacetime without gravity \cite{tHooft,Susskind,RMP}.
The fundamental assertion of the holographic principle is that the number of degrees of freedom (DoF) in the bulk space is given by the amount of information relatively stored in the quantum pure/purified state of the boundary quantum field theory.

The $d+2$-dimensional anti-de Sitter spacetime/$d+1$-dimensional conformal field theory (AdS$_{d+2}$/CFT$_{d+1}$) correspondence is the known example of the holographic principle \cite{AdSCFT1,GKP,W,AdSCFT2}.
In the context of the AdS$_3$/CFT$_2$ correspondence, the author argued in Ref.\cite{Konishi1} that the holographic principle asserts that the anti-de Sitter bulk space has {\it negative} (i.e., stochastic) classicalized spin DoF dual to the boundary quantum pure/purified state after the classicalization of this boundary quantum state.

Here, in the language of the ensemble interpretation of quantum mechanics \cite{dEspagnat}, {\it classicalization} means reducing the quantum pure ensemble of a state vector, which consists of qubits, to the {\it classical} mixed ensemble of product qubit eigenstates (where {\it classical} means there is no quantum interference of qubits) \cite{Konishi2}.
Thus, after classicalization, entropy is generated (i.e., information is lost).

Without respect to the holographic principle, negative DoF exist in quantum theory as its probabilistic nature.
For example, the quantum mechanical expectation value $\la \widehat{q}^a\ra$ of a particle position $q^a$ at a given time $t$ consists of
(I) the explicit {\it positive} (i.e., deterministic) orbital DoF $(q^a_t,\dot{q}^a_t)$ in the Lagrangian of the particle and
(II) the implicit negative DoF in the operation of taking the expectation value by using a wave function.

If all orbital DoF are originally {\it only} positive DoF, in quantum theory we can ask, what is the origin of the negative DoF?
In this article, we attempt to derive these negative DoF from the classicalized holographic tensor network (cHTN), which itself has {\it only} negative classicalized spin DoF.
Here, HTN indicates bulk-space realization of the multi-scale entanglement renormalization ansatz (MERA) of a boundary quantum pure/purified state \cite{Vidal1,Vidal2} in the context of the AdS$_3$/CFT$_2$ correspondence \cite{Swingle1,Matsueda,Review,Tak,Swingle2}.
At each cHTN site (defined later), a positive/negative classicalized spin DoF is expressed as a statistical mixture of two of four eigenstates of bipartite qubits.
This expression is thus bivalent.

Specifically, we derive the imaginary-time path-integral of a non-relativistic particle \cite{Book1} from the action of the cHTN given in Ref.\cite{Konishi1}, assuming that this particle originally has no negative DoF.
Schematically, we state that
\begin{eqnarray}
&&{\rm positive\ DoF}+{\rm gravity}\ ({\rm negative\ DoF})\nonumber\\
&&={\rm path\ integral\ of\ positive\ DoF}\;,
\end{eqnarray}
and this scheme gives rise to the quantum mechanics of the positive orbital DoF by taking the inverse Wick rotation from (real-valued) imaginary time to (real-valued) real time.

Here, note that there are as many negative classicalized spin DoF in the cHTN as there are sites in the cHTN.
Each site with a classicalized entangled bipartite-qubit state is the minimum unit of the negative curvature in the cHTN (i.e., the anti-de Sitter space).
These negative classicalized spin DoF are thus those of {\it gravity} in the cHTN.

The key step in this derivation of the imaginary-time path-integral is to reinterpret the Euclidean action of a free particle as the action of positive classicalized spin DoF.
Then, we obtain a classical stochastic process of the particle in imaginary time.
In the presence of potential energy, we need to consider particle annihilation by the potential energy in imaginary time.
To treat this effect, we enlarge the original ensemble of events for a free particle with the null ensemble.

The rest of this article is organized as follows.
In Sec. II, in preliminaries, we introduce and explain the basic ideas of the HTN, that is, the MERA, and its classicalization.
Section III presents the main results of this article.
Specifically, we derive the imaginary-time path-integral of a non-relativistic particle from the holographic principle.
In Sec. IV, we present an interpretation and translations of three concepts.
In Sec. V, we summarize the overall results of this article.

Throughout this article, we choose $(-,+,+)$ as the signature of the AdS$_3$ spacetime metric after the inverse Wick rotation, and hatted symbols are operators.

\section{Preliminaries}
The purpose of this section is to introduce the cHTN of the HTN, which is dual to the ground state of a strongly coupled CFT$_2$.

A {\it tensor network} is a one-extradimensional geometrization of quantum entanglements in a quantum many-body pure state, which consists of qubits \cite{Swingle2}.
A {\it tensor} is a multilinear operator that acts on a state vector of a qubit or qubits \cite{Swingle2}.
Here, quantum entanglements are generated by the interactions between qubits.
So, as a quantum many-body system is strongly coupled, its ground state is heavily entangled.

To treat heavy quantum entanglements in a quantum many-body pure state such as the ground state $|\psi\ra$ at a quantum critical point (i.e., in conformal field theory), the MERA was introduced as a scale-invariant tensor network model.
The scale invariance of the MERA reflects the scale invariance in conformal field theory.
The MERA consists of two types of operations in the entanglement renormalizations, namely, the {\it coarse-grainer} (an isometry) $\widehat{W}$ and the {\it disentangler} $\widehat{U}$ \cite{Vidal1,Vidal2}.
Intuitively, a coarse-grainer replaces bipartite qubits with a coarse-grained unipartite qubit, and a disentangler resolves an entangled bipartite-qubit state into a bipartite-qubit product state locally.

Here, the coarse-grainer $\widehat{W}$ maps a bipartite-qubit state in the four-dimensional state space $V \otimes V$ to a coarse-grained unipartite-qubit state in the two-dimensional state space $V^{\prime}$.
Then, $\widehat{W}$ satisfies the conditions
\begin{equation}
\widehat{W}^\dagger\widehat{W} =\widehat{\Pi}\;,\ \ \widehat{W} \widehat{W}^\dagger=\widehat{1}_{V^\prime}\;,
\end{equation}
where the dagger represents the Hermitian conjugate, and $\widehat{\Pi}$ is the projection operator onto a two-dimensional linear subspace of $V\otimes V$, satisfying $\widehat{\Pi}^2=\widehat{\Pi}$.
Next, the disentangler $\widehat{U}$ is an automorphism in $V\otimes V$ and satisfies unitarity
\begin{equation}
\widehat{U}^\dagger \widehat{U}=\widehat{U}\widehat{U}^\dagger=\widehat{1}_{V\otimes V}\;.
\end{equation}
The MERA is the real-space renormalization group transformation of the quantum pure state $|\psi\ra$ in the boundary CFT$_2$ by the semi-infinitely alternate combinations of the layer of the disentanglers and the layer of the coarse-grainers \cite{Vidal1,Vidal2}.
For our own terminology, we define the {\it site} of the MERA by the four qubits involved with a disentangler $\widehat{U}$, and the {\it entangled bipartite-qubit state} at a site means the bipartite-qubit state entangled by the entangler $\widehat{U}^\dagger$.

Now, the MERA (hereinafter, we call it the HTN) is a quantum pure state.
Traditionally, to quantify quantum entanglements in the HTN, we separate the HTN into two parts, whose inner part is the so-called {\it timeslice of entanglement wedge}, by a Ryu--Takayanagi curve (a geodesic) \cite{RT1,RT2} in the HTN and trace out the outer part in this separation.
Then, we obtain the entanglement entropy as the von Neumann entropy of this reduced {\it quantum} mixed state of the inner part of the HTN.
As holographic duality, this entanglement entropy is that of the subregion state, dual to the entanglement wedge in the HTN, in the boundary CFT$_2$.

The cHTN is a classical mixed state with respect to the qubits.
To obtain it, we decohere the quantum pure state of the HTN in the eigenbasis of the qubits in the HTN completely.
This means that we restrict the set of observables to the set of the ones that commute with the third Pauli matrix, which is diagonal in the qubit eigenbasis.
Namely, we introduce a {\it superselection rule} (i.e., selection of the quantum mechanical observables which commute with a given superselection operator) in the set of observables.
Here, the superselection operator for the one-qubit Hilbert space is the third Pauli matrix.

In the cHTN at the strong-coupling limit of CFT$_2$, the von Neumann entropy of this classical mixed state is given by the discretized area of the cHTN \cite{Konishi1,Konishi3}.
Entropy is the lost information.
Then, the HTN itself has no information, but the cHTN itself has negative information.
The holographic principle suggests that this negative information gives rise to negative DoF of classicalized {\it spins} in the cHTN as the anti-de Sitter bulk space.

\section{Derivation}

The action of the cHTN is that of classicalized spins with negative DoF \cite{Konishi1}
\begin{equation}
I_{\rm bulk}[|\psi\ra]=-\hbar b H_{\rm bdy}^{\rm bit}[|\psi\ra]\label{eq:original}
\end{equation}
for the bit factor $b=\ln 2$ and the boundary quantum ground state $|\psi\ra$.
$H_{\rm bdy}^{\rm bit}[|\psi\ra]$ is the {\it measurement entropy} (i.e., the von Neumann entropy of the classical mixed state obtained by classicalization) of $|\psi\ra$ in bits.
According to the holographic principle, the negative amount of information $-H_{\rm bdy}^{\rm bit}[|\psi\ra]$ in bits is the negative number of DoF of classicalized spins in the cHTN.
Note that this action is time independent.

We consider the action, which is the sum of the action of the cHTN with negative classicalized spin DoF and the (Euclidean) action, that is, the imaginary-time integral of the imaginary-time Hamiltonian $S[\gamma_\tau]$ of a non-relativistic free particle
\begin{equation}
I_{\rm bulk}[|\psi\ra,\gamma_\tau]=-\hbar b H_{\rm bdy}^{\rm bit}[|\psi\ra]+S[\gamma_\tau]\label{eq:original2}
\end{equation}
for a spatial path $\gamma_\tau$ with specified edge points in the anti-de Sitter bulk space, parametrized by the imaginary time $\tau$ (s.t., $0\le \tau\le 1$).

At the strong-coupling limit of the boundary CFT$_2$, the measurement entropy $H_{\rm bdy}^{\rm bit}[|\psi\ra]$ is maximized with respect to $|\psi\ra$, and this action is evaluated as \cite{Konishi1,Konishi3}
\begin{equation}
I_{\rm bulk}[\gamma_\tau]=-\hbar b A_{\rm TN}+S[\gamma_\tau]\label{eq:actionI}
\end{equation}
for the discretized area $A_{\rm TN}$ of the cHTN.
This evaluated action can be rewritten as
\begin{equation}
I_{\rm bulk}[\gamma_\tau]=-\hbar b \biggl(A_{\rm TN}-b^{-1}\frac{S[\gamma_\tau]}{\hbar}\biggr)\;,\label{eq:action}
\end{equation}
where the first term means that the cHTN has a local branch (i.e., a classicalized entangled bipartite-qubit state) into {\it two} spin eigenstates in a classical mixed state at each site, and this local branch loses {\it one bit} of information per site.
From the second term, therefore, we can interpret $b^{-1}S[\gamma_\tau]/\hbar$ for each path $\gamma_\tau$ as {\it information} in bits (i.e., a series of spin eigenstate selections at local branches in the cHTN) by noting the negative sign in this second term.
For this information in nats, we can define its number of series of spin eigenstates (events) $W_{\gamma_\tau}^{\rm spin}$ for each path $\gamma_\tau$ and the classical probability $p^{\rm cl}_{\gamma_\tau}$ of obtaining a series of spin eigenstates along $\gamma_\tau$ as \footnote{Note that these definitions can be used only when all events have equal probability.
Eq.(\ref{eq:actionI}) satisfies this condition.}
\begin{equation}
\frac{S[\gamma_\tau]}{\hbar}=\ln W_{\gamma_\tau}^{\rm spin}=-\ln p^{\rm cl}_{\gamma_\tau}\;.
\end{equation}
From this, we obtain
\begin{equation}
p^{\rm cl}_{\gamma_\tau}=\exp\biggl({-\frac{S[\gamma_\tau]}{\hbar}}\biggr)\;.\label{eq:prob}
\end{equation}
The events of this probability are the series of spin eigenstates, one of which accompanies a path $\gamma_\tau$.
When we consider the conditional probability density $P(q_1^a,1|q_0^a,0)$ of the particle being found at a point in space $q_1^a=\gamma_1^a$ with $\tau=1$ conditioned by another point in space $q_0^a=\gamma_0^a$ with $\tau=0$ for $a=1,2$, then from Eq.(\ref{eq:action}) a generic event of this conditional probability density is
\begin{equation}
(q_1^a,1)|(q_0^a,0)\simeq \coprod_{\gamma_\tau}(\gamma_\tau|{\rm cHTN})\bigr|_{\gamma_0^a=q_0^a}^{\gamma_1^a=q_1^a}\;,\label{eq:stoc}
\end{equation}
where the coproduct (exclusive union) with respect to the probe $\gamma_\tau$ is due to its arbitrariness in Eq.(\ref{eq:action}) and its classicality.
This conditional probability density is
\begin{eqnarray}
P(q_1^a,1|q_0^a,0)&=&\int_{\gamma_0^a=q_0^a}^{\gamma_1^a=q_1^a} {\cal D}{\gamma_\tau} p^{\rm cl}_{\gamma_\tau}\label{eq:condP}\\
&=&\int_{\gamma_0^a=q_0^a}^{\gamma_1^a=q_1^a} {\cal D}{\gamma_\tau} \exp\biggl({-\frac{S[\gamma_\tau]}{\hbar}}\biggr)
\end{eqnarray}
for the Wiener measure ${\cal D}\gamma_\tau$ of the path $\gamma_\tau$.
This is the path-integral representation of the density matrix of a free particle in the equilibrium quantum statistical mechanics up to a normalization constant (i.e., the partition function with inverse temperature of unity) \cite{Book1}:
\begin{equation}
P(q_1^a,1|q_0^a,0)=\sum_n \phi_n(q_1^a)\bar{\phi}_n(q_0^a)\exp\biggl({-\frac{\Delta\tau}{\hbar}E_n}\biggr)\label{eq:proof}
\end{equation}
for $\Delta \tau\equiv 1$, and the $n$-th eigenfunction $\phi_n(q^a)$ and the $n$-th eigenvalue $E_n$ of the Hamiltonian of the particle.

In the absence of potential energy,
\begin{equation}
\int dq_1^aP(q_1^a,1|q_0^a,0)=1\label{eq:total}
\end{equation}
holds.
This is because, in this case, the equilibrium quantum statistical mechanics is equivalent to a classical stochastic process in imaginary time.
On the other hand, in the presence of potential energy in the equilibrium quantum statistical mechanics, the particle can be annihilated in the classical stochastic process in imaginary time, and Eq.(\ref{eq:total}) does not hold.\footnote{Free classical Brownian motion in a medium is an analogy to this classical stochastic process.
See Ref.\cite{PR}.}
Here, this particle annihilation by the potential energy in imaginary time is dual to the particle scattering by the potential energy in real time.
Then, the original ensemble ${\cal E}$ of events for the particle in the absence of potential energy would be enlarged to a statistical mixture with the null ensemble $\emptyset$, and the potential-energy part of $S[\gamma_\tau]/\hbar$ would be the {\it information} of the original ensemble ${\cal E}$ in nats in the enlarged ensemble comprising the original ensemble ${\cal E}$ and the null ensemble $\emptyset$ (see Fig.1).

\begin{figure}[htbp]
\begin{center}
\includegraphics[width=0.4\hsize,bb=2 2 257 257]{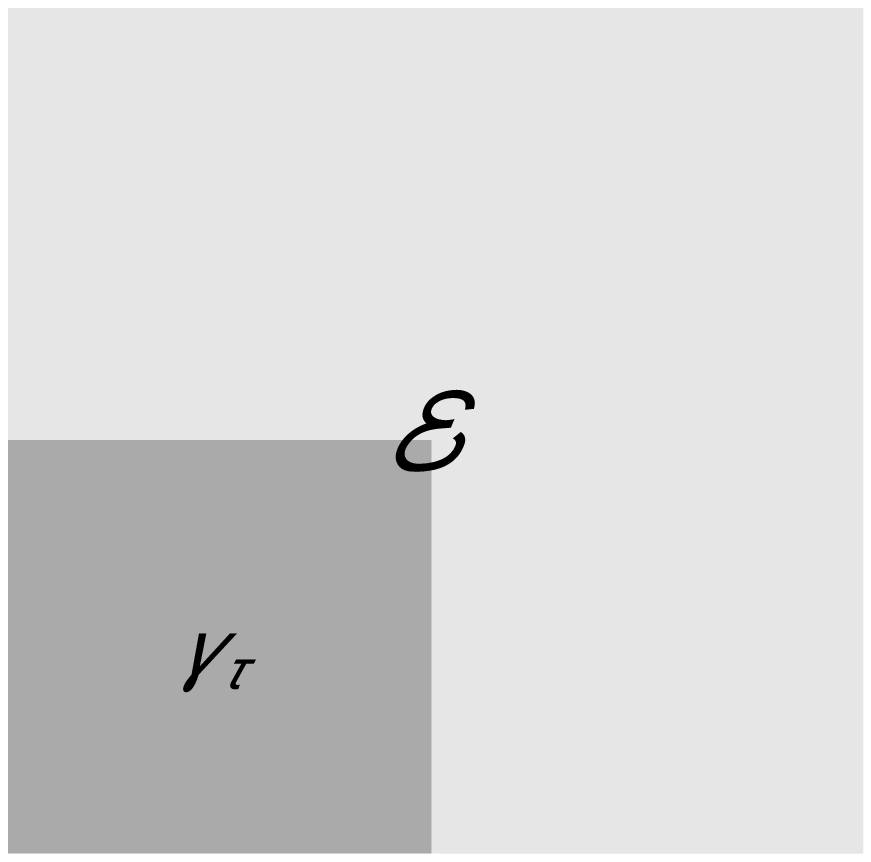}
\includegraphics[width=0.4\hsize,bb=2 2 257 257]{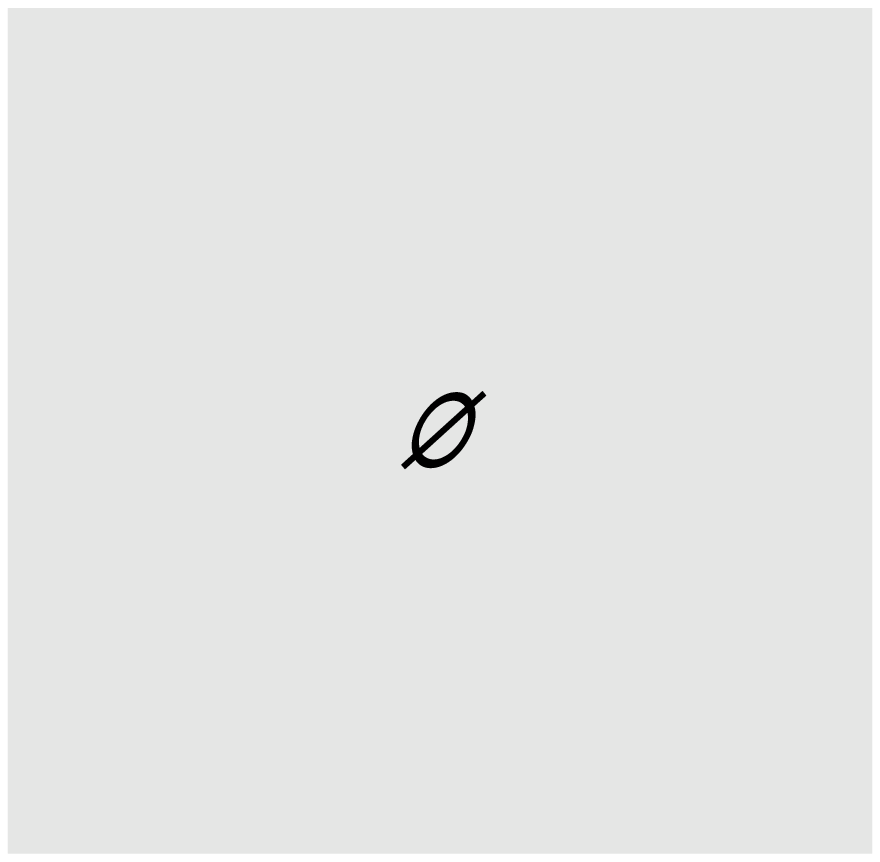}
\includegraphics[width=0.4\hsize,bb=2 2 257 257]{Ensemble1Thick.eps}
\includegraphics[width=0.4\hsize,bb=2 2 257 257]{Ensemble2Thick.eps}
\end{center}
\caption{Schematic of the enlarged ensemble in the presence of potential energy.
$\emptyset$, ${\cal E}$, and $\gamma_\tau$ represent the null ensemble, the original ensemble, and a path event in the original ensemble, respectively.
The potential-energy part of $e^{-S[\gamma_\tau]/\hbar}$ is the ratio of the original ensemble ${\cal E}$ in the enlarged ensemble $\{{\cal E},{\cal E},\emptyset,\emptyset\}$.
Here, it is $1/2$.
$e^{-S[\gamma_\tau]/\hbar}$ in the absence of potential energy is the ratio of a path event $\gamma_\tau$ in the original ensemble ${\cal E}$.
Here, it is $1/4$.
$e^{-S[\gamma_\tau]/\hbar}$ in the presence of potential energy is the product of these two ratios.
Here, it is $1/8$.}
\end{figure}

Now, we introduce the real time $t$ by taking the inverse Wick rotation of the imaginary time $\tau$ as
\begin{equation}
t=-i \tau\;.\label{eq:Wick}
\end{equation}
Then, we obtain a quantity
\begin{equation}
K(q_1^a,1|q_0^a,0)=\int_{\gamma_0^{\ast a}=q_0^a}^{\gamma_1^{\ast a}=q_1^a} {\cal D}{\gamma^\ast_t} \exp \biggl({i\frac{S^\ast[\gamma^\ast_t]}{\hbar}}\biggr)\label{eq:A}
\end{equation}
from Eq.(\ref{eq:condP}).
For Eq.(\ref{eq:proof}), this quantity is the quantum probability amplitude of a free particle being found at a spacetime point $(q_1^a,1)$ conditioned by another spacetime point $(q_0^a,0)$:
\begin{equation}
K(q_1^a,1|q_0^a,0)=\sum_n\phi_n(q_1^a)\bar{\phi}_n(q_0^a)\exp\biggl({-\frac{i\Delta t}{\hbar}E_n}\biggr)
\end{equation}
for $\Delta t\equiv 1$.
This is a formal and consistent correspondence between the equilibrium quantum statistical mechanics in imaginary time and quantum mechanics in real time \cite{Book1}.
In Eq.(\ref{eq:A}), $S^\ast$ is the Lorentzian action of the particle obtained from its action $S$ by the inverse Wick rotation (\ref{eq:Wick}), $\gamma^\ast_t$ with specified edge points $q_0^a=\gamma^{\ast a}_0$ and $q_1^a=\gamma^{\ast a}_1$ is a path in the anti-de Sitter bulk spacetime, parametrized by the real time $t$ (s.t., $0\le t\le 1$), and ${\cal D}\gamma^\ast_t$ is the Wiener measure of the path $\gamma^\ast_t$.
In contrast to the conditional probability density (\ref{eq:condP}), the absolute square of the quantum probability amplitude (\ref{eq:A}) represents the classical probability density of an event {\it only when} quantum measurement of the event is performed.

Here, we elaborate on the inverse Wick rotation (\ref{eq:Wick}) in our context.
Before the inverse Wick rotation (\ref{eq:Wick}), events in the {\it Euclidean} bulk spacetime (meaning the metric signature $(+,+,+)$) are {\it off-shell} and are given by the increments of the Euclidean action $S[\gamma_\tau]$ of a particle by $\hbar b$ along an off-shell imaginary-time path $\gamma_\tau$ with specified edge points, which is an event in itself and always exclusively summed with other off-shell imaginary-time paths (the other events) in a statistical mixture (\ref{eq:stoc}) (i.e., occurs objectively).
For a free particle, these events originate from the quantum mechanical events in the classical mixed state of $|\psi\ra$ in the boundary CFT$_2$.
After the inverse Wick rotation (\ref{eq:Wick}), events in the {\it Minkowskian} bulk spacetime (meaning the metric signature $(-,+,+)$) are {\it on-shell} in their expectation values and are the quantum mechanical and relativistic events obtained by quantum measurements.
The inverse Wick rotation (\ref{eq:Wick}) from the imaginary time $\tau$ to the real time $t$ changes the causal structure of events in the bulk spacetime from the Newtonian {\it total ordering} of off-shell events to the relativistic {\it partial ordering} of events of local observables.
In the absence of quantum measurements (i.e., the absence of quantum mechanical events), the unitary quantum mechanical process of a particle after the inverse Wick rotation (\ref{eq:Wick}) is dual to the classical stochastic process of the particle before the inverse Wick rotation (\ref{eq:Wick}): this is a dimensional descent of the source of events.

\section{Interpretation and translations}

In this section, we present an interpretation and translations of three concepts.

First, the meaning of the classical probability $p^{\rm cl}_{\gamma_\tau}$ for a free particle that will not be annihilated in imaginary time is as follows.
For the action $S[\gamma_\tau]$ reinterpreted as that of the positive classicalized {\it spin} DoF with the amount of $b^{-1}S[\gamma_\tau]/\hbar$, bipartite qubits are in their eigenstates.
On the other hand, the cHTN is a classical mixed state of bipartite qubits (i.e., one has a negative classicalized spin DoF) at each cHTN site.
Their spin eigenstates are thus selected from the classical mixed state at sites, and their probability is $1/2$, because at each HTN site there are the Bell state entanglements of bipartite qubits when the measurement entropy is maximized (i.e., the strong-coupling limit of the boundary CFT$_2$ is taken: the anti-de Sitter bulk spacetime is purely general relativistic) by the most probable configuration of the cHTN.
Then, we obtain the classical probability $p^{\rm cl}_{\gamma_\tau}$ in Eq.(\ref{eq:prob}) by multiplying this probability $1/2$ over all classicalized spins at sites with the number of $b^{-1}S[\gamma_\tau]/\hbar$.
Namely, both $W_{\gamma_\tau}^{\rm spin}$ and $p^{\rm cl}_{\gamma_\tau}$ decompose into products of quantities with respect to a single spin eigenstate in times of $b^{-1}S[\gamma_\tau]/\hbar$.

Next, we translate this imaginary-time bulk process before the inverse Wick rotation (\ref{eq:Wick}) into the boundary language.
By taking the classicalization of the HTN, the quantum pure state $|\psi\ra$ in CFT$_2$ on the boundary is completely decohered in the eigenbasis of the qubits.
At this time, information in bits with the amount of $H_{\rm bdy}^{\rm bit}[|\psi\ra]$ relatively stored in the quantum coherence of $|\psi\ra$ is completely lost \cite{Konishi1,Konishi3}.
On the boundary side, the addition of the term $S[\gamma_\tau]$ to the original action (\ref{eq:original}) means that, for an off-shell imaginary-time path $\gamma_\tau$, lost information in bits with the amount of $b^{-1}S[\gamma_\tau]/\hbar$ is acquired in total by a series of alternative selections of eigenstates in the classicalized boundary states of the entanglement wedges in the cHTN.
Contrary to the quantum measurement, each classicalized boundary state itself does not collapse to a classical pure state after these alternative selections of eigenstates.
This is because these selections are attributed to the term $S[\gamma_\tau]$ in the action (\ref{eq:actionI}).

Finally, without respect to the holographic principle, we translate the orbital superselection rule, which plays an indispensable role in the quantum measurement processes (i.e., decoherence, and readout of an event) proposed in Refs.\cite{Araki,Konishi4}, in the quantum mechanics of a non-relativistic system with two orbital DoF (e.g., a non-relativistic abstracted {\it macroscopic} measurement apparatus) in the anti-de Sitter bulk spacetime after the inverse Wick rotation (\ref{eq:Wick}) into the classical stochastic process in imaginary time before the inverse Wick rotation (\ref{eq:Wick}).
We introduce the orbital superselection rule by discretizing (redefining) the position $q^a$ and the momentum $p^a$ with their partition widths $\epsilon_{Q^a}$ and $\epsilon_{P^a}$ (s.t., $\epsilon_{Q^a}\epsilon_{P^a}=h$), respectively, so that all redefined orbital observables commute with each other owing to the von Neumann theorem proved in Ref.\cite{Neumann}.
Then, the original Hilbert space of the state vectors of the system is based by the simultaneous eigenstates $\{|Q^a,P^a\ra\}$ of the redefined position $\widehat{Q}^a$ and the redefined momentum $\widehat{P}^a$ \cite{Neumann}, and we obtain
\begin{eqnarray}
&&\la Q_1^a,P_1^a|\widehat{O}|Q_0^a,P_0^a\ra \nonumber \\
&&=\delta_{Q_0^a,Q_1^a}\delta_{P_0^a,P_1^a}\la Q_0^a,P_0^a|\widehat{O}|Q_0^a,P_0^a\ra\label{eq:SSR}
\end{eqnarray}
for an arbitrary redefined orbital observable $\widehat{O}$ of the system, which commutes with all redefined orbital observables of the system.
Namely, in the presence of the orbital superselection rule, a given quantum pure state of the system automatically becomes equivalent to a classical mixed state of simultaneous eigenstates of the redefined position and the redefined momentum with respect to all redefined orbital observables of the system.
In Eq.(\ref{eq:SSR}), because $Q^a$ and $P^a$ are discrete observables, their Kronecker deltas appear instead of their Dirac delta functions.
When we choose $\widehat{O}$ in Eq.(\ref{eq:SSR}) as $\exp(-i\widehat{H}/\hbar)$ for the redefined Hamiltonian $\widehat{H}$ of the system, we obtain
\begin{eqnarray}
&&K((Q_1^a,P_1^a),1|(Q_0^a,P_0^a),0)\nonumber\\
&&=\delta_{Q_0^a,Q_1^a}\delta_{P_0^a,P_1^a}K((Q_0^a,P_0^a),1|(Q_0^a,P_0^a),0)\;.
\end{eqnarray}
Then, by taking the Wick rotation (\ref{eq:Wick}), we obtain
\begin{eqnarray}
&&P((Q_1^a,P_1^a),1|(Q_0^a,P_0^a),0)\nonumber\\
&&=\delta_{Q_0^a,Q_1^a}\delta_{P_0^a,P_1^a}P((Q_0^a,P_0^a),1|(Q_0^a,P_0^a),0)\;.
\end{eqnarray}
Because this formula holds not only for $\Delta \tau=1$ but also for an arbitrary value of $\Delta \tau$, the orbital superselection rule applied for a simultaneous eigenstate $|Q_0^a,P_0^a \ra$ of the system replaces the off-shell imaginary-time paths of the system with the given Planck cell (i.e., an event) $(Q_0^a,P_0^a)$ in the phase space.
Because of Eq.(\ref{eq:SSR}), starting from a generic quantum pure state of the system, the orbital superselection rule gives rise to a statistical mixture of Planck cells (i.e., events) $\{(Q_0^a,P_0^a)\}$ in the phase space with respect to all redefined orbital observables of the system.

\section{Summary}

In this article, in the context of the AdS$_3$/CFT$_2$ correspondence, we attempted to derive the quantum mechanics of the positive DoF of a non-relativistic particle in the anti-de Sitter bulk spacetime from the holographic principle.
Here, we give an overview of our main results as shown in Fig.2.
In our derivation, we start from the quantum pure state $|\psi\ra$ of the ground state of the strongly-coupled boundary CFT$_2$, which is dual to the quantum pure state of the HTN.
Next, we completely decohere the HTN to the cHTN in the eigenbasis of the qubits.
Then, the cHTN is a classical mixed state with respect to the qubits, and information stored in the HTN is completely lost.
From the holographic principle and the results in Refs.\cite{Konishi1,Konishi3} about the amount of lost information, the action of the cHTN is that of negative classicalized spin DoF.
For the positive DoF of a {\it free} particle in the cHTN, their Euclidean action in imaginary time is reinterpreted as that of a series of positive classicalized spin DoF.
One positive classicalized spin DoF arises when the action of the positive DoF increases by $\hbar b$ along an off-shell imaginary-time path.
Then, each generation of the positive classicalized spin DoF means the alternative selection of a spin eigenstate at a site of the cHTN.
As the imaginary-time path of the particle in the cHTN is exclusively arbitrary and classical, the positive DoF are always in a statistical mixture of classical pure states (i.e., a classical mixed state) of off-shell imaginary-time paths.
In the presence of potential energy, we incorporate particle annihilation by the potential energy in imaginary time into this classical mixed state by enlarging the original ensemble of events for a free particle with the null ensemble (see Fig.1).
Finally, by taking the inverse Wick rotation from the imaginary time to the real time, we obtain from this classical mixed state in imaginary time a quantum pure state (i.e., quantum mechanics) of the positive DoF in the real-time description.

\begin{figure}[htbp]
\begin{center}
\includegraphics[width=0.8\hsize,bb=2 2 257 257]{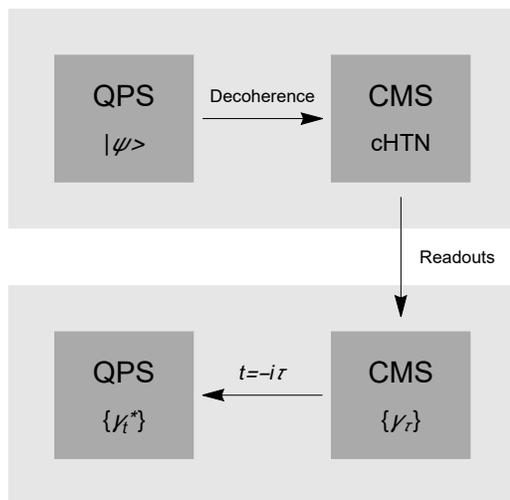}
\end{center}
\caption{Schematic of our derivation of the quantum mechanics of positive DoF from the holographic principle.
In this figure, QPS and CMS are abbreviations for quantum pure state and classical mixed state, respectively.}
\end{figure}

In addition to this derivation of quantum mechanics, we showed that the orbital superselection rule in a non-relativistic quantum mechanical system with two orbital DoF replaces the off-shell paths of this system in real/imaginary time with a statistical mixture of Planck cells (i.e., events) in the phase space with respect to the orbital observables of this system which are selected by the orbital superselection rule.


\clearpage

\begin{thebibliography}{99}
\bibitem{tHooft}G. 't Hooft,
%%"Dimensional reduction in quantum gravity",
arXiv:gr-qc/9310026.
\bibitem{Susskind}L. Susskind,
%%"The world as a hologram",
J. Math. Phys. {\bf 36}, 6377 (1995).
\bibitem{RMP}R. Bousso,
%%"The holographic principle", 
Rev. Mod. Phys. {\bf 74}, 825 (2002).
\bibitem{AdSCFT1}J. M. Maldacena, 
%%"The large-N limit of superconformal field theories and supergravity", 
Adv. Theor. Math. Phys. {\bf 2}, 231 (1998).
\bibitem{GKP}S. S. Gubser, I. R. Klebanov and A. M. Polyakov,
%%"Gauge theory correlators from non-critical string theory",
Phys. Lett. B {\bf 428}, 105 (1998).
\bibitem{W}E. Witten,
%%"Anti de Sitter space and holography",
Adv. Theor. Math. Phys. {\bf 2}, 253 (1998).
\bibitem{AdSCFT2}O. Aharony, S. S. Gubser, J. M. Maldacena, H. Ooguri and Y. Oz,
%% "Large-N field theories, string theory and gravity",
 Phys. Rep. {\bf 323}, 183 (2000).
\bibitem{Konishi1}E. Konishi,
%% "Addendum: Holographic interpretation of Shannon entropy of coherence of quantum pure states",
 EPL {\bf 132}, 59901 (2020), arXiv:1903.11244.
\bibitem{dEspagnat}B. d'Espagnat, {\it Conceptual Foundations of Quantum Mechanics,} 2nd edn. (W. A. Benjamin, Reading, Massachusetts, 1976).
\bibitem{Konishi2}E. Konishi,
%% "Random walk of bipartite spins in a classicalized holographic tensor network",
 Results in Physics {\bf 19}, 103410 (2020).
\bibitem{Vidal1}G. Vidal,
%% "Entanglement renormalization",
 Phys. Rev. Lett. {\bf 99}, 220405 (2007).
\bibitem{Vidal2}G. Vidal, 
%%"Class of quantum many-body states that can be efficiently simulated", 
Phys. Rev. Lett. {\bf 101}, 110501 (2008).
\bibitem{Swingle1}B. Swingle, 
%%"Entanglement renormalization and holography", 
Phys. Rev. D {\bf 86}, 065007 (2012).
\bibitem{Matsueda}H. Matsueda, M. Ishibashi and Y. Hashizume,
%%"Tensor network and a black hole",
Phys. Rev. D {\bf 87}, 066002 (2013).
\bibitem{Review}N. Bao, C. Cao, S. M. Carroll, A. Chatwin-Davies and N. Hunter-Jones,
%%"Consistency conditions for an AdS multiscale entanglement renormalization ansatz correspondence",
Phys. Rev. D {\bf 91}, 125036 (2015).
\bibitem{Tak}M. Rangamani and T. Takayanagi, Lect. Notes Phys. {\bf 931}, 1 (2017).
\bibitem{Swingle2}B. Swingle, 
%%"Spacetime from entanglement",
 Annu. Rev. Condens. Matter Phys. {\bf 9}, 345 (2018).
\bibitem{Book1}R. P. Feynman and A. R. Hibbs, {\it Quantum Mechanics and Path Integrals} (McGraw-Hill, New York, 1965).
\bibitem{RT1}S. Ryu and T. Takayanagi, 
%%{Holographic derivation of entanglement entropy from the anti-de Sitter space/conformal field theory correspondence},
Phys. Rev. Lett. {\bf 96}, 181602 (2006).
\bibitem{RT2}S. Ryu and T. Takayanagi, 
%%{Aspects of holographic entanglement entropy}, 
J. High Energy Phys. {\bf 08}, 045 (2006).
\bibitem{Konishi3}E. Konishi, 
%%"Holographic interpretation of Shannon entropy of coherence of quantum pure states", 
EPL {\bf 129}, 11006 (2020).
\bibitem{PR}F. W. Wiegel, 
%%"Path integral methods in statistical mechanics", 
Phys. Rep. {\bf 16}, 57 (1975), pp. 59-60.
\bibitem{Araki}H. Araki, {Prog. Theor. Phys.} {\bf 64}, 719 (1980).
\bibitem{Konishi4}E. Konishi, arXiv:2012.01886.
\bibitem{Neumann}J. von Neumann, {\it Mathematical Foundations of Quantum Mechanics} (Princeton University Press, Princeton, NJ, 1955).
\end{thebibliography}
\end{document}